\documentclass[fleqn,twoside,twocolumn,nofootinbib]{revtex4} % Specifies the document class %,unsortedaddress
\usepackage[sec]{ujp} % \usepackage[cyr]{ujp} for cyrillic, \usepackage[web]{ujp} for web
\begin{document}
\title[POWER SPECTRUM OF RADIATION FROM A GAUSSIAN SOURCE]%колонтитул
{POWER SPECTRUM OF RADIATION FROM A GAUSSIAN SOURCE
MICROLENSED BY A POINT MASS:\\ ANALYTIC RESULTS}%
\author{V.I.~Zhdanov}%1 автор
\affiliation{Taras Shevchenko National University of Kyiv, Astronomical Observatory}%институт
\address{3, Observatorna Str., Kyiv 04053, Ukraine}%адрес
\email{ValeryZhdanov@gmail.com}%e-mail
\affiliation{National Technical University of Ukraine ``Kyiv
Polytechnic Institute''}%
\address{37, Prosp. Peremogy, Kyiv 03056, Ukraine}%
\author{D.V.~Gorpinchenko}
\affiliation{National Technical University of Ukraine ``Kyiv
Polytechnic Institute''}%
\address{37, Prosp. Peremogy, Kyiv 03056, Ukraine}%
\udk{524.8}  \razd{\secxi}

\setcounter{page}{1083}%
\maketitle

\begin{abstract}
Gravitational lensing deals with general-relativistic effects in
the propagation of electromagnetic radiation. We consider
wavelength-dependent contributions in case of a (micro)lensing of
an extended Gaussian source by a point mass under standard
assumptions about the incoherent emission of different source
elements.
 Analytical expressions for the power spectrum of
 a microlensed radiation, which are effective
 in  case of a large source, are obtained. If the
source center, the mass, and an observer are on a straight line,
the power spectrum is found in a closed form in terms of a
hypergeometric function. In the case of general locations of the lens
and the source, the result is presented in the form of a series.
Approximate analytic expressions for the power spectrum in the case of
a large source and high frequencies are obtained.
\end{abstract}

\section{Introduction}
Effects of the propagation of electromagnetic radiation in a curved
space-time, which is described within the General Relativity, is
 referred to as the gravitational lensing. This direction
 has an important meaning for investigations of
mass distributions in the Universe, for probing the planetary and
solar-mass  objects in the Milky Way and for investigations of
the innermost structures of quasars \cite{Wambs, Tisserand, OGLE,
SchEhlFal, KochSchneiWambs, AlSlZh}. One of the most vivid examples
of gravitational lensing applications is the confirmation of the
dark matter existence in the Bullet Cluster \cite{direct_DM_2}. The
term ``microlensing'' is used when the radiation from a remote
source is deflected by the gravitational fields of stellar and/or
planetary mass objects \cite{Wambs, SchEhlFal, KochSchneiWambs}.

All the effects observed so far in the gravitational lens systems
(GLSs) are well described within geometric optics approximations,
none of these problems deals with the wave optics. The detection
of wavelength-dependent effects in GLSs could open the way to
entirely new tests of electrodynamics in a curved space-time. On
the other hand, the effects of the physical optics
 might give an independent information
about the lensing objects. Related problems have been studied
theoretically  starting from the early papers \cite{Mandzhos_1981,
SchSchmB, Deguchi}; for the later progress see \cite{Mandzhos_1991a,
Mandzhos_1991b, Mandzhos_1995, jaroszynski, zakhar,
zabel,Matsunaga}; a complete  bibliography can be found in
monographs \cite{Bliokh, SchEhlFal, MinakovVaculik}.

All the papers deal with different theoretical aspects of the wave
optics in weak gravitational fields; there are no observational
results. The reason is that the wave effects in GLSs are either very
small or very improbable. Nevertheless, Heyl
\cite{Heyl_1,Heyl_2} turned attention recently that the  wave optics
effects having a gravitational origin can be measured in some
future in the case of the microlensing by planetary masses and
planetesimals. A number of authors (see, e.g.,
\cite{Matsunaga,takahashi} and references therein) study
diffractive effects in GLSs in connection with the expected
detection of gravitational waves.

Observational signatures of the microlensed radiation from  real sources are
blurred because of coherence properties; as those in the usual optics,
they strongly depend on the size of an extended source. The
diffractive microlensing of extended sources has been studied in
 caustic crossing events \cite{jaroszynski, zabel} and
in the case of point mass lensing  \cite{Matsunaga}; the mutual
coherence of different lensed images of an extended source near
caustics has been estimated in a series of papers by Mandzhos
\cite{Mandzhos_1991a, Mandzhos_1991b, Mandzhos_1995}. It should be
pointed out that, in GLSs, the diffraction effects go side by side with
the interference between different images of every source point. The
occurrence of different images leads to additional maxima or
deformations of the autocorrelation function of the microlensed
radiation (see, e.g., \cite{Borra1997, Borra2008, Borra2011,
Zhdanov}). Both diffraction and interference in GLSs  can be
considered on an equal basis by means of a power spectrum of the
microlensed radiation.

Most researches on this problem involve numerical calculations.
Nevertheless, it is desirable to have  analytic results at least
for some simple problems. In this paper, we found such a result in
the case of a Gaussian source; this result has been unknown earlier in
spite of rather a long history of investigations in this field.
We derive the power spectrum of the radiation from a Gaussian
source, which is microlensed by one point mass under standard
assumptions about
 incoherent source elements. In Section \ref{basicrelations},
 the radiation field is obtained in a
standard way using the Kirchhoff integral. In Section \ref{central}
and Appendix A, we derive a closed relation in terms of a
hypergeometric function for the power spectrum, when the lensing
mass is projected onto the center of the source.  In Section
\ref{noncentral}, we use this relation in the case of  a general
  source disposition to obtain approximations for
the power spectrum for a sufficiently small mass and a large source.
As distinct from the earlier method used in this problem
\cite{Matsunaga}, our analytic approach is convenient in the case of
a sufficiently large source size as compared to the Einstein
radius projected onto the source plane. We also present the first
terms of an expansion in powers of $1/\omega$ in the
high-frequency case.

\section{Basic relations}\label{basicrelations}

In this section, we formulate common relations used below. Following
\cite{Deguchi, Matsunaga} and many other authors, we use standard
considerations of the diffraction theory and the gravitational
lensing (see, e.g., \cite{Bliokh, SchEhlFal}); respectively, we
neglect the polarization. The calculations of the field are performed in the
flat space-time background. This is relevant, e.g., in the case of the
 Milky Way systems; however, the results can be easily
extended to the case of extragalactic GLSs after some redefinition
of distances in a curved space-time.

Leaving aside the polarization effects, we describe the radiation
field by means of one scalar function $\varphi(t,{\bf r})$.
Furthermore, we work in the Cartesian coordinates; the observer, the
lens, and the source center are situated in a neighborhood of
the $Z$-axis in the planes $z = 0,\,z = D_d$ and $z = D_s,$ respectively.
As we neglect the polarization, we describe the source ``current'' with a
scalar function ${\rm { j}}(t,{\rm {\bf y}}),$ by assuming that this
source lies completely in the plane ${\rm {\bf r}} = ({\rm {\bf
y}},D_s),\,{\rm {\bf y}}\in {\bf R}^2$. We also assume that ${\rm {
j}}(t,{\rm {\bf y}})$ is a stochastic process having the correlation
properties
%1
 \begin{equation}
\label{averj}
 \langle{\rm {j}}(t,{\rm
{\bf r}}){\rm  j}(t',{\rm {\bf r}}') \rangle = \delta ({\rm {\bf y}}
- {\rm {\bf y}}')I(t - t',{\rm {\bf y}});
\end{equation}

\noindent this relation presumes that different points of the source
are incoherent; $\langle \ldots \rangle$ represent an ensemble
average.

For the Fourier transform  $ {\rm {  \tilde {j}}}(k,{\rm {\bf y}}) =
\frac{1}{\sqrt {2\pi } }\int_{ - \infty }^\infty {\rm { j}}(t,{\rm
{\bf y}})\times$ $\times e^{i\omega t}dt,$  we have
%2
\begin{equation}
 \label{favery}
 \langle {\rm {  \tilde {j}}}(\omega ,{\rm {\bf y}}){\rm {  \tilde {j}}}^\ast
(\omega ',{\rm {\bf y}}')\rangle = \delta ({\rm {\bf y}} - {\rm {\bf
y}}')\delta (\omega - \omega ')\,\tilde {I}(\omega ,{\rm {\bf y}}),
\end{equation}
where $\tilde {I}(\omega ,{\rm {\bf y}}) = \int {dt} \,I(t,{\rm {\bf
y}})e^{i\omega t}$  is an intensity  of an element at a point ${\rm
{\bf y}}$ for the frequency $\omega $.

The Helmholtz equation for $\tilde \varphi(\omega,{\bf r})$ is
\[
\Delta \tilde \varphi + \omega^2 \tilde \varphi= -4\pi \tilde j
\quad (c = 1).
\]
\*Consider a solution $\tilde \varphi(\omega,{\bf r})$ of this
equation describing the radiation from the source plane. Let $D_{ds}
= D_s - D_d$ be the distance from the lens plane to the source
plane; the source is situated near the $Z$-axis: $|{\bf y} |\ll D_{ds}
$. Before the lens plane ${\rm {\bf r'}} = ({\rm {\bf y'}} ,D_d+0)$
near the $Z$-axis ($ |{\bf y'} |\ll D_{ds}$), the solution is obtained as that
in the flat-space diffraction theory \cite{LL,wavetheory}:
%3
 \[ \tilde {\varphi
}_b(\omega ,{\rm {\bf y'}}) =\]
\begin{equation}
\label{sollenspl}= \frac{e^{i\omega D_{ds} }}{D_{ds} }\int {d^2} {
{\bf {y}}} \,\,{\rm { \tilde {j}}}(\omega ,{\rm {\bf {y}}} )\exp
\left[ {\frac{i\omega }{2D_{ds} }({\rm {\bf y'}} - {\rm {\bf {y}}}
)^2} \right].
\end{equation}
In order to calculate the field after passing the lens plane, we use
the well-known approach of the phase screen (see, e.g.,
\cite{Bliokh, SchEhlFal, MinakovVaculik}), by assuming that the
radiation gains an additional phase shift $\omega t_{\rm grav}$ on
the lens plane, where $t_{\rm grav} ({\rm {\bf y'}}) $ is the
general relativistic time delay for signals crossing the lens plane
at ${\bf y}'$. This delay is the same for all frequencies. Then the
solution just after passing the lens plane reads
\[
\tilde {\varphi}_a (\omega ,{\rm {\bf y}}') =e^{-i\omega t_{\rm
grav}({\bf y'}) }\tilde {\varphi }_b (\omega ,{\rm {\bf y}}').
\]
The radiation comes to an observer at the origin  ${\rm {\bf r}}
\bf{=0}$. The field $\tilde {\varphi }(\omega ,{\rm {\bf 0}})$  is
calculated by means of the Kirchhoff--Sommerfeld method
\cite{wavetheory}:
%4
\[\tilde {\varphi }(\omega ,{\rm {\bf 0}}) = \frac{\omega
e^{i\omega D_d }}{2\pi i{\kern 1pt} D_d }\int {d^2} {\rm {\bf
y}}'\tilde {\varphi }_a (\omega ,{\rm {\bf y}}')\exp \left[
{\frac{i\omega }{2D_d }{\rm {\bf y}}^{\prime 2}} \right]=\,\]
\begin{equation}
\label{obsfield}=
 \frac{e^{i\omega D_s }}{2  i{\kern 1pt} D_d
D_{ds} } \int {d^2} {\rm {\bf y}}\,{\rm { \tilde {j}}}(\omega ,{\rm
{\bf y}}) \phi (\omega ,{\rm {\bf y}}),
\end{equation}

\noindent where
%5
\[ \phi (\omega ,{\rm {\bf y}}) =\]
\begin{equation}\label{phi} =
\frac{\omega }{\pi }\int {\exp \left\{ {i\omega \left[ {\frac{({\rm
{\bf x}} - {\rm {\bf y}})^2}{2D_{ds} } + \frac{{\rm {\bf x}}^2}{2D_d
} - t_{\rm grav} ({\rm {\bf x}})} \right]} \right\}} d^2{\rm {\bf
x}} \,.
\end{equation}

Using (\ref{favery}, \ref{obsfield}), we obtain
\[
 \langle \tilde {\varphi }(\omega ,{\rm {\bf 0}})\tilde {\varphi }^\ast ({\omega
}',{\rm {\bf 0}}) \rangle = \delta (\omega - {\omega }')P(\omega )
\]
\noindent with the power spectrum
%6
\begin{equation}
\label{power}P(\omega ) = \left( {\frac{1}{2{\kern 1pt} D_d D_{ds}
}} \right)^2\int {d^2} {\rm {\bf y}}\,\tilde {I}(\omega ,{\rm {\bf
y}})\left| {\phi (\omega ,{\rm {\bf y}})} \right|^2.
\end{equation}

\section{Central Gaussian Source and Point Lensing Mass}\label{central}

Relation (\ref{power}) is written for arbitrary sources and
gravitational time delays. Further, we consider the case of one point
microlens at ${\rm {\bf r}} = ({\bf 0},D_d )$ with the delay time
(e.g., \cite{Bliokh, SchEhlFal, MinakovVaculik})
%7
\begin{equation}
 \label{tgrav}
 t_{\rm grav} ({\rm {\bf y}}) = 2r_g\ln (\vert {\rm {\bf
y}}\vert / L),\quad r_g=2Gm ;
\end{equation}

\noindent here, $m$ is the microlens mass, $L$ is a dimensional
parameter which disappears in final calculations; further, it is
omitted.

We assume the Gaussian brightness distribution over a source for
the intensity ${\tilde I}$ from Eq. (\ref{favery}):
%8
\begin{equation}
\label{eq1} I(\omega ,{\rm {\bf y}},{\rm {\bf r}}_0 ) =
\frac{f(\omega )}{\pi R^2}\exp \left( { - \frac{({\rm {\bf y}} -
{\rm {\bf r}}_0 )^2}{R^2}} \right).
\end{equation}
\noindent Here, ${\rm {\bf r}}_0 $ is the source center in the
source plane, the function $f(\omega)$ is supposed to be the same for
all source points and determines the coherence properties
of emitting source elements.

Integral (\ref{phi})  can be calculated exactly
\cite{Deguchi,Bliokh,Matsunaga} in terms of the confluent
hypergeometric function  $\Phi (a,c;x)$ \cite{Bateman}:
 %9
 \[\phi (\omega ,{\rm {\bf y}}) =\]
\[=\frac{\omega }{\pi }\int {\exp \left\{ {i\omega \left[
{\frac{({\rm {\bf x}} - {\rm {\bf y}})^2}{2D_{ds} } + \frac{{\rm
{\bf x}}^2}{2D_d } - 2r_g \ln \left| {\rm {\bf x}} \right|)}
\right]} \right\}} d^2{\rm {\bf x}}=
\]
\[
 = \omega ^{i\sigma }e^{\frac{i\omega {\rm {\bf y}}^2}{2D_s }}\Gamma (1 -
i\sigma )\left( {\frac{2D_{ds} D_d i}{D_s }} \right)^{1 - i\sigma
}\times\]
\begin{equation}\label{eq3}
\times  \Phi \left( {\;i\sigma ,1;\;i\sigma y^2 / R_{E,s}^2 }
\right),
\end{equation}
\noindent where we used the Kummer transformation \cite{Bateman};
$\sigma = \omega r_g ,\, R_{E,s} =[ 2r_g D^{*}]^{1/2} $ is the
Einstein radius projected onto the source plane, and $D^{*}=D_{ds} D_s
/ D_d $; $y = \left| {\rm {\bf y}} \right|$.

For the Gaussian source, the power spectrum (\ref{power}) is
as follows:
%10
\[ P(\omega ,{\rm {\bf r}}_0 ) = \left( {\frac{1}{2{\kern 1pt} D_d
D_{ds} }} \right)^2\frac{f(\omega )}{\pi R^2}\times
\]
\begin{equation} \label{eq4}
\times  \int {d^{\,2}} {\rm {\bf y}}\,\exp \left[ { - \frac{({\rm
{\bf y}} - {\rm {\bf r}}_0 )^2}{R^2}} \right]\left| {\phi (\omega
,{\rm {\bf y}})} \right|^2.
\end{equation}
The microlensing effect can be described by the ratio
%11
\begin{equation}
\label{upsilon} \Upsilon = P(\omega ,{\rm {\bf r}}_0 ) / P_0 (\omega
),
\end{equation}
where $P_0$ is the power spectrum in the absence of the microlensing
($\sigma = 0)$.

If the source center is at the origin (${\rm {\bf r}}_0 = 0)$,
integral (\ref{eq4}) can be estimated in terms of hypergeometric
functions \cite{Vovkogon}; the derivation (see Appendix A) yields
$\Upsilon = \Upsilon _0 (\alpha ,\sigma ),$ where
%12
 \[
\Upsilon _0 (\alpha ,\sigma ) \equiv \exp [2\sigma \arctan(\beta
)]\left| {\Gamma (1 - i\sigma )} \right|^2 \times\]
\begin{equation}
\label{eq5}\times F\left( {i\sigma , - i\sigma ;\,1\,;(1 + \beta ^2
)^{ - 1}} \right);
\end{equation}

\noindent  $\beta=\alpha/\sigma,\,\alpha = R_{E,s}^2 / R^2$,
$R>0$; and $F(a,b;c;x)$ is the hypergeometric function. Note that
$\sqrt{\beta}=R^{-1}\sqrt{\lambda D^*/\pi}$ plays the role of
the ratio of the Fresnel zone size to the source size.

For large $\alpha \gg 1$ and  $\sigma\sim O(1),$ the argument of the
hypergeometric function is small, and we have $ \Upsilon _0 (\alpha
,\sigma ) \approx 2\pi \sigma$.

For  $\alpha \ll 1$, i.e., $R_{E,s} \ll R$, and bounded $\sigma
\sim O( 1),$ it is convenient to use the expansion of the
hypergeometric function $F(a,b;c;x)$ near the point $x = 1$ for an
integer  $c$. Then formula (\ref{eq5}) takes the form
%13
\[\Upsilon _0
(\alpha ,\sigma ) = \exp [2\sigma \arctan(\beta )] \times\]
\begin{equation}
\label{eq6}\left\{ {1 - \sigma ^2s\sum\limits_{n = 0}^\infty
{\left| {\frac{\Gamma (n + 1 + i\sigma )}{\Gamma (1 + i\sigma )}}
\right|^2\frac{s^n[k_n(\sigma)- \ln s]}{(n + 1)(n!)^2}} }
\right\},
\end{equation}
where
\[  k_n(\sigma)= 2\psi (n + 1)  - \psi (n + 1 + i\sigma ) - \psi
(n + 1 - i\sigma )+\] \[+ \frac{1}{n + 1}, \]
\[  s = \alpha ^2 / (\alpha ^2 + \sigma ^2), \quad \psi (x) \equiv
{d\ln \Gamma (x)}/{dx}.\]
We write the expansion in $\alpha $
up to the terms $\sim \alpha ^2$ and $\alpha ^2\ln \alpha $ that
depend on the frequency. In this approximation, Eq.
(\ref{eq6}) can be written as
%14
\begin{equation}
\label{eq7} \Upsilon _0 (\alpha ,\sigma ) = 1 + 2\alpha + 2\alpha
^2 - \alpha ^2[k_0 - 2\ln (\alpha / \sigma )] \quad .
\end{equation}
Expansion (\ref{eq6}) can be rewritten in the form, which is
convenient to look for the asymptotic expansions at large frequencies:
%15
\[ \Upsilon_0 (\alpha ,\sigma ) = \exp [2\sigma \arctan (\alpha /
\sigma )]\times \]
\begin{equation}
\label{expans}\times\left\{ {1 - \sum\limits_{n = 0}^\infty {C_n
(\sigma )\frac{\tilde {s}^{n + 1}[\tilde {k}_n(\sigma) - \ln \tilde
{s}]}{(n + 1)(n!)^2}} } \right\},
\end{equation}
where $ \tilde {s} = \sigma ^2s = \alpha ^2/(1 +
\beta ^2)$,
%16
\[
\tilde k_n(\sigma)= k_n(\sigma)+2ln(\sigma),\]
\begin{equation}
\label{tildas} C_n (\sigma ) = \left| {\frac{\Gamma (n + 1 + i\sigma
)}{\Gamma (1 + i\sigma )\sigma ^n}} \right|^2 = \prod\limits_{m =
0}^n {\left( {1 + \frac{m^2}{\sigma ^2}} \right)}.
\end{equation}
Using the asymptotic relations for the function $\psi $ (e.g.,
\cite{Bateman}) at large arguments, the coefficients $\tilde
k_n(\sigma)$ can be expanded in powers of $\sigma^{-2}$. Therefore,
we can write
%17
\begin{equation}
\label{asymptexp}
 \Upsilon _0 (\alpha ,\sigma ) = \sum\limits_{n =
0}^m {\sigma ^{ - 2n}\Upsilon _0^{(n)} (\alpha )} + O(\sigma ^{ -
2(m + 1)}).
\end{equation}
Up to the terms $\sim \sigma^{-2},$ we have
%18
 \begin{equation}
 \tilde {k}_n (\sigma )=  \tilde{k}_n (\infty ) - \frac{1}{\sigma ^2}\left[ {n(n + 1) +
\frac{1}{6}} \right] + O(\sigma ^{ - 4}),
\end{equation}
$ \tilde {k}_n (\infty ) = 2\psi (n + 1) + (n + 1)^{-1}.$
It follows from (\ref{tildas}) that
\[
C_n (\sigma )  = 1 + \frac{n(n + 1)(2n + 1)}{6\,\sigma ^2} + ...
\]
Then we have the geometric optics limit ($\sigma\equiv\omega r_g
\to \infty$)
%19
\[ \Upsilon _0 (\alpha ,\infty )= \Upsilon _0^{(0)}
(\alpha ) = \]
 \[= \exp (2\alpha )\left\{ {1 - \alpha ^2\sum\limits_{n
= 0}^\infty {\frac{\alpha ^{2n}[\tilde {k}_n (\infty ) - 2\ln \alpha
]}{(n + 1)(n!)^2}} } \right\}=\]
\begin{equation}
\label{geolim}=2\alpha e^{2 \alpha } K_1(2\alpha )\,,
\end{equation}
where we used a representation for the modified Bessel function $K_1$
(see formula 8.446 in \cite{gradshtein}). A direct calculation within
geometric optics (Appendix B) is in accordance with this expression.

For the next term of the expansion, we have
\[
 \Upsilon _0^{(1)} (\alpha ) = - \alpha ^2e^{2\alpha }\left\{
 \frac{2}{3}\alpha + \sum\limits_{n = 0}^\infty {\frac{C_n^{(1)}(\alpha)\alpha
^{2n}}{(n + 1)(n!)^2}}\right\},
\]
\[
C_n^{(1)}(\alpha)=  \alpha^2 - n(n + 1) - \frac{1}{6} +\]\[+\left(
\tilde {k}_n (\infty ) - 2\ln \alpha \right)  \left[ \frac{n}{6}(n +
1)(2n + 1) - \alpha ^2(n + 1) \right] \,.
\]

\section{Non-Central Source}\label{noncentral}
 We now proceed to the derivation of $\Upsilon$ in the case of a non-central source, by
 using Eqs. (\ref{eq5})--(\ref{eq7}). The integration over the angular variable
yields
%20
 \[ P(\omega ,{\rm {\bf r}}_0 )
= \left( {\frac{1}{2{\kern 1pt} D_d D_{ds} }}
\right)^2\frac{2f(\omega )}{R^2}
 \exp \left( { - \frac{ {\rm {\bf r}}_0^2}{R^2}} \right)\times\]
 \begin{equation}
\label{eq8}\times \int\limits_0^\infty {dy} \,y\exp \left( { -
\frac{y^2}{R^2}} \right)I_0 \left( {\frac{2yr_0 }{R^2}}
\right)\left| {\phi (\omega ,{\rm {\bf y}})} \right|^2,
\end{equation}
\noindent where $r_0 = \left| {{\rm {\bf r}}_0 } \right|$. Using the
expansion of the modified Bessel function $I_0 $ in a Taylor series
and the substitution  $y^2 = tR_{E,s}^2,$ we get
%21
\[ \Upsilon (\alpha ,\sigma ,r_0 )=
\alpha \exp \left( { - \frac{r_0^2 }{R^2} + \pi \sigma }
\right)\left| {\Gamma (1 - i\sigma )} \right|^2 \times \]
\begin{equation} \label{eq9} \times \sum\limits_{n = 0}^\infty
{\frac{\alpha ^n}{(n!)^2}\left( {\frac{r_0 }{R}} \right)}
^{2n}\int\limits_0^\infty {dt\;} e^{ - \alpha t}t^n\left| {\Phi
\left( {\;i\sigma ,1;\;i\sigma t} \right)} \right|^2.
\end{equation}
Using the differentiation with respect to the parameter $\alpha$ and
taking Eqs. (\ref{eq3}) and (\ref{eq5}) into account, we
obtain
%22
\[ \Upsilon (\alpha ,\sigma ,r{ }_0)=\exp \left( {
- \frac{r_0^2 }{R^2}} \right)\times \]
\begin{equation} \label{eq10} \times \sum\limits_{n = 0}^\infty {( - 1)^n\frac{\alpha ^{n +
1}}{(n!)^2}\left( {\frac{r_0 }{R}} \right)}
^{2n}\frac{\partial^n}{\partial\alpha ^n}\left[ {\frac{\Upsilon _0
(\alpha ,\sigma )}{\alpha }} \right].
\end{equation}
Obviously, this representation is workable when ${r_0}/{R}$ is not
large; this will be supposed further. This excludes, e.g., the case
of a point source.

The simple analysis shows that, to obtain the terms up to the orders
of $\alpha ^2$ and $\alpha ^2\ln \alpha $  in (\ref{eq10}), it
is sufficient to use the truncated expression
 (\ref{eq7}). We get
 %23
\[
 \Upsilon (\alpha ,\sigma ,r_0 ) = 1 + 2\alpha e^{ - r_0^2 / R^2}
 +\alpha ^2e^{ - r_0^2 / R^2} \left\{ 2g\left( \frac{r_0^2
 }{R^2}\right)-\right.
 \]
\begin{equation}
\label{noncentralappro}\left.-2\frac{r_0^2 }{R^2}+
  \left( 1 - \frac{r_0^2 }{R^2}\right)\left[ 2 - k_0 (\sigma ) - 2\ln (\sigma / \alpha )
  \right]
\right\},
 \end{equation}
\noindent where
\[
 g(x) = \sum\limits_{n = 2}^\infty
{\frac{x^n}{n{\kern 1pt} (n - 1){\kern 1pt} n!}} =
\]
\[= (x -
1)[Ei(x) - C - \ln x] - e^x + 1 + 2x,
\]
$Ei(x)$ is the integral exponent, and $C$ is the Euler constant.

We now obtain the asymptotic expansion for $\Upsilon (\alpha
,\sigma ,r_0 )$ at large frequencies. The substitution of
(\ref{asymptexp}) into (\ref{eq10}) leads to the asymptotic series
in powers of $\sigma^{-2}$:
%24
\begin{equation}\label{asympt1} \Upsilon (\alpha ,\sigma ,r_0 ) = \sum\limits_{m}^M
{\sigma ^{ - 2m}\Upsilon ^{(m)}(\alpha ,r_0 )} + O(\sigma ^{ - 2(M +
1)})
\end{equation}
with the coefficients
%25
\[ \Upsilon ^{(m)}(\alpha ,r_0 ) = \exp \left( { -
\frac{r_0^2 }{R^2}} \right)\times \]
\begin{equation}\label{coeffs}\times\sum\limits_{n = 0}^\infty {( -
1)^n\frac{\alpha ^{n + 1}}{(n!)^2}\left( {\frac{r_0 }{R}} \right)}
^{2n}\frac{\partial ^n}{\partial \alpha ^n}\left[ {\frac{\Upsilon
_0^{(m)} (\alpha )}{\alpha }} \right].
\end{equation}
Some caution is needed when we differentiate the asymptotic expansion
(\ref{asymptexp}) term by term; this can be easily justified in view
of the explicit structure of the functions involved. The geometric
optics limit yields
%26
\begin{equation}\label{geomopt1}
 \Upsilon (\alpha ,\infty ,r_0 ) = \Upsilon
^{(0)}(\alpha ,r_0 ),
\end{equation}
where $\Upsilon ^{(0)}(\alpha ,r_0 )$ is given by (\ref{coeffs}) for
$m=0$. The direct calculation of the amplification using the standard
gravitational lensing theory yields the same result (see Appendix
B).

The coefficient $\Upsilon ^{(1)} (\alpha )$, which describes the
first correction to (\ref{geomopt1}) with an accuracy  up to $\sim
\alpha ^2$, can be obtained either from (\ref{coeffs}) or from
(\ref{noncentralappro}):
%27
\begin{equation}
\label{noncentrhighf}\Upsilon  ^{(1)} (\alpha ) = \frac{1}{6}\alpha
^2\exp \left( { - \frac{r_0^2 }{R^2}} \right)\left( {1 - \frac{r_0^2
}{R^2}} \right).
\end{equation}

\section{Discussion}

We have obtained the analytic expressions for the radiation power
spectrum for an  extended Gaussian source microlensed by a point
mass under standard assumptions about the incoherence of different
source elements. Our results allow one to treat the cases of wide
and narrow bandwidth receptions on the same basis by an appropriate
choice of the function $f(\omega)$ in Eq. (\ref{eq1}). If the source
center, the lensing mass, and the observer are situated on one
straight line, the power spectrum is  given by Eq. (\ref{eq5}) in
terms of a hypergeometric function. In the case of a general
arrangement, the result is presented in the form of the functional
series (\ref{eq10}). This representation of the power spectrum
enables us to derive approximations up to any accuracy in the case
of sufficiently small $\alpha=(R_{E,s} / R)^2$. The representation
is efficient under the condition that the distance $r_0$ between the
lens projection onto the source plane and the source center is
comparable/less than the source size. The opposite case ($r_0\gg R$)
can be treated by the method developed in \cite{Matsunaga}, where
the expansion of the magnification around the source position is
used. Representations (\ref{eq5}) and (\ref{eq10}) have been used to
derive the lowest orders of the asymptotic expressions
(\ref{noncentralappro}) and (\ref{noncentrhighf}) in the cases of a
small lens
 and high frequencies. We have shown by means of a direct calculation that the high
frequency limit yields exactly the expressions of the geometric
optics. As we see from (\ref{eq7}) and (\ref{noncentrhighf}), the
first nontrivial wave optics contribution, which is dependent on the
frequency, appears in the terms $\sim \alpha^2$ (though the first
geometric optics lensing contribution has order $\sim \alpha$); it
disappears for very large sources. At high frequencies, this
contribution behaves itself as $\sim \omega^{-2}$.\looseness=1

In what follows, we propose some estimates for the radio waveband;
though much smaller wavelengths also can be of interest (cf., e.g.,
\cite*{stanek}). For the effects of the wave optics to be
significant, one must have $\sigma=r_g\omega\sim 1$ and $\alpha \sim
1$. For a wavelength $\sim $~1--10~cm, the first condition is
fulfilled if the microlens mass is of the order of $10^{-5}
M_{\odot}$. Such planetary mass objects must be common in the Milky
Way, and there is a number of observational confirmations of
exosolar planets, including microlensing observations (e.g.,
\cite{Wambs}). Heyl \cite{Heyl_1} pointed out that the signatures of
the diffractive gravitational microlensing can be detected during
the occultation of distant stars by Kuiper-Belt and Oort cloud
objects. However, in this case, the treatment must be modified by
the introduction of an opaque screen that describes the lensing
object (cf. \cite{Bliokh}).

In order that $\alpha$ be not small, the source size $R$ must be of
the order or less than $R_{E,s}$. This is true for a wide interval
of $D^*$ in the case of the microlensing by the Milky Way objects;
however, in this case, the probability of the microlensing of a
suitable radio source is small (see, however, \cite{Heyl_2}). For a
distant extragalactic source at $D_s\approx D_{ds}\sim 10^{3}$~Mpc
microlensed by a planet at $D_d=10$~kpc, we have $R_{E,s}\sim
10^{-2}[M/(10^{-5} M_{\odot})]^{1/2}$~pc. The typical size of
extragalactic radio sources  is larger, though it may have
inhomogeneous structures with typical scales $\sim R_{E,s}$. In this
case, the wavelength-dependent effects of gravitational lensing will
be noticeable.

\vskip3mm This work has been supported in part by the
``Cosmomicrophysics'' program of National Academy of Sciences of
Ukraine and SCOPES grant №128040 of Swiss National Science
Foundation.

\subsubsection*{APPENDIX A.\\ Derivation of Eq.
(\ref{eq5}) in the case of a central source }\label{AA}

{\footnotesize From Eq. (10), we have
%28
\[
 \left| {\phi (\omega ,{\rm {\bf y}})} \right|^2 = \left|
{\Gamma (1 - i\sigma )} \right|^2\left( {\frac{2D_{ds} D_d }{D_s }}
\right)^2 \times\]
\begin{equation}
 \label{A1}\times e^{\sigma \pi }\left| {\Phi \left(
{\;i\sigma ,1;\;i\sigma y^2 / R_{E,s}^2 } \right)} \right|^2 .
\end{equation}

For $r_0=0,$ the power spectrum (\ref{eq4}) equals
%29
\begin{equation}
\label{eq4zero}  P(\omega , 0 ) =  \left( {\frac{1}{2{\kern 1pt} D_d
D_{ds} }} \right)^2\frac{2f(\omega )}{ R^2}\int\limits_0^\infty d {
y}\,y\exp \left( { - \frac{{  { y}} ^2}{R^2}} \right)\left| {\phi
(\omega ,{ { y}})} \right|^2,
\end{equation}
where the integration over the angular variable is performed. With
regard for (\ref{A1}) and (\ref{eq4zero}) and by using the
substitution $t = y^2$, we obtain ratio (\ref{upsilon})
corresponding to ${\rm {\bf r}}_0 = 0$:
%30
\[ \Upsilon _0 (\alpha ,\sigma ) =\]
\begin{equation}
 \label{A2}=
e^{\sigma \pi }\left| {\Gamma (1 - i\sigma )} \right|^2\int {dt}
\,e^{ - t}\left| {\Phi \left( {i\sigma ,1;\;i\sigma t / \alpha }
\right)} \right|^2 \, ,
\end{equation}
where $\alpha = R_{E,i}^2 / R^2.$ \*The integral in formula
(\ref{A2}) is a special case of the expression that can be obtained
by means of formula 6.15.22 in \cite{Bateman}. Here, in order
to estimate (\ref{A2}), we calculate directly
the integral
%31
\begin{equation}
\label{AA} I(a,\lambda ,a\,',{\lambda }') = \int\limits_0^\infty
{dt} \,e^{ - t}\Phi \left( {a,1;\;\lambda t} \right)\Phi \left(
a\,',1;\;{\lambda }'t \right),
\end{equation}
where $0<$ Re$(a) < 1$, $0<$ Re$(a\,')  < 1,\;\left| {\lambda }'
\right| + \left| \lambda \right| < 1$.
 With this aim, we need representations for the confluent hypergeometric function
  %32
\begin{equation}
 \label{A4}
 \Phi \left( {a,1;\;x} \right) =\frac{1}{B(a,1-a)} \int\limits_0^1
{du} \;\frac{e^{xu}u^{a - 1}}{(1 - u)^a}
\end{equation}
and for the hypergeometric function
%33
\[ F\left( {a,b;1;\;x} \right) = \frac{1}{\Gamma (b)\Gamma (1 -
b)}\int\limits_0^1 {du} \;\frac{t^{b - 1}}{(1 - zt)^a(1 - t)^b} =\]
\begin{equation}
 \label{A3}= \frac{1}{B(b,1 - b)}\int\limits_0^\infty {d\tau }
\;\frac{\tau ^{b - 1}(1 + \tau )^{a - 1}}{\left[ {1 + (1 - z)\,\tau
} \right]^{a}} \, ,
\end{equation}
 \noindent where $B(a,b)$
is the Beta-function, and $t = \tau / (1 + \tau )$.

We use (\ref{A4}) for $\Phi \left( {a,1;\;\lambda t} \right)$ and
$\Phi \left( {{a}',1;\;{\lambda }'t} \right)$ in (\ref{AA}) and
integrate over $t$ after the change of the integration
order:
\[
I(a,\lambda ,a\,',{\lambda }') = \frac{1}{B(a,1 - a) B(a\,',1 -
a\,')} \times\]\[ \times \int\limits_0^1 {du} \;u^{a - 1}(1 - u)^{ -
a}\int\limits_0^1 {dv} \;\frac{v^{{a}' - 1}(1 - v)^{ - {a}'}}{1 -
\lambda u - {\lambda }'v}.
\]
\noindent The integration over $dv$ after the substitution $v
\rightarrow \xi$ ($v = \tau / (1 + \tau )$ and  $\tau = \xi (1 -
\lambda u)/(1 - \lambda - \lambda u)$) is easily carried out with
the use of an integral representation for the Beta-function.

Then the substitution $u = \eta / (1 -\lambda + \eta )$  yields
\[
I(a,\lambda ,a\,',{\lambda}') =\]\[ = \frac{(1 - \lambda )^{ - a}(1
- {\lambda }')^{ - a\,'}}{B(a,1 - a)}\,\int\limits_0^\infty {d\eta }
\;\frac{{\eta}^{a - 1}(1 + \eta )^{a\,' - 1}}{\left[ {1 + (1 - z)
\eta } \right]^{a\,'}},
\]
where $   z =  \lambda {\lambda }'(1 - \lambda )^{-1}(1 - {\lambda
}')^{-1}$. Therefore, in view of representation (\ref{A3}),
we get the final relation
%34
\begin{equation} \label{A5}
I(a,\lambda ,a\,',{\lambda}') =  \frac{F({a}\,',a;\;1;\;z)}{(1 -
\lambda )^a(1 - {\lambda }')^{{a}'}}.
\end{equation}

All calculations are fulfilled in the domain of the parameters
$a,\lambda ,{a}',{\lambda }'$, where the integrals involved are
convergent. However, relation (\ref{A5}) can be analytically
continued to a wider domain, which includes the values $a =
i\sigma ,\;{a}' = - i\sigma ,\;\lambda = i\sigma / \alpha
,\;{\lambda }' = - i\sigma / \alpha$. This yields Eq.
(\ref{eq5}) of the main text.

}

\subsubsection*{APPENDIX B.\\ Gaussian source amplification in
geometric optics }

{\footnotesize The amplification of a point source by a point mass
lens within the geometric optics in variables of the source plane is
well known \cite{SchEhlFal,Bliokh}:
%35
\begin{equation}
\label{copoint} \mu(y)=\frac{y^2+2 R_{E,s}^2}{y\sqrt{y^2+4
R_{E,s}^2}}.
\end{equation}
 In our
case, this must be convolved with the Gaussian brightness
distribution (\ref{eq1}) yielding the total amplification
\[
A =\frac{1}{\pi R^2}\int d^{\,2}{\bf y} \mu(y) e^{-({\bf y}-{\bf
r_0})^2/R^2}=\]
\[= \frac{2}{R^2}e^{- \frac{ r_0^2}{R^2}} \int
\limits_0^\infty dy\,y \,\mu(y) e^{- \frac{ y^2}{R^2}}
I_0\left(\frac{2yr_0}{R^2}.\right)
\]
Taking (\ref{copoint}) into account and using the Taylor expansion
of $I_0$ and the substitution $y^2=t R_{E,s}^2,$ we have
%36
\[A=A(\alpha,r_0)=\]\[=e^{- \frac{ r_0^2}{R^2}}\sum\limits_{n=0}^\infty
\frac{\alpha^{n+1}}{(n!)^2}\left(\frac{ r_0}{R }\right)^{2n} \int
\limits_0^\infty dy\,  e^{- \alpha t} t^{n}\frac{t+2}{\sqrt{t^2+4
t}}=\]
\begin{equation} \label{cogauss} = e^{- \frac{
r_0^2}{R^2}}\sum\limits_{n=0}^\infty
\frac{(-1)^n\alpha^{n+1}}{(n!)^2}\left(\frac{ r_0}{R }\right)^{2n}
\frac{\partial^n}{\partial \alpha^n}\int \limits_0^\infty dy\, e^{-
\alpha t}  \frac{t+2}{\sqrt{t^2+4 t}} .
\end{equation}
The latter integral can be written \cite{gradshtein} in terms of the
modified Bessel function $K_1$:
 %37
\begin{equation}
\label{cogauss2} A(\alpha, 0)=\alpha \int \limits_0^\infty dt\, e^{-
\alpha t}
 \frac{t+2}{\sqrt{t^2+4 t}} =2\alpha e^{2\alpha}K_1(2\alpha).
 \end{equation}
This is the same as $\Upsilon _0 (\alpha ,\infty )$ of
(\ref{geolim}).

For an arbitrary source position, relation (\ref{cogauss}) yields
 %38
\begin{equation}
\label{cogauss3} A(\alpha,r_0)=e^{- \frac{
r_0^2}{R^2}}\sum\limits_{n=0}^\infty
\frac{(-1)^n\alpha^{n+1}}{(n!)^2}\left(\frac{ r_0}{R }\right)^{2n}
\frac{\partial^n}{\partial \alpha^n}
\left[\frac{A(\alpha,0)}{\alpha}\right] ,
\end{equation}
which is the same
as $\Upsilon (\alpha ,\infty,r_0 )$ of (\ref{geomopt1}).

}

\rezume{СПЕКТР ПОТУЖНОСТІ ВИПРОМІНЮВАННЯ \\ВІД ГАУСІВСЬКОГО ДЖЕРЕЛА,
МІКРОЛІНЗОВАНОГО\\ ТОЧКОВОЮ~~~~ МАСОЮ:~~~~ АНАЛІТИЧНІ~~~~
РЕЗУЛЬТАТИ}{В.І. Жданов, Д.В. Горпінченко}{Теорія гравітаційного
лінзування вивчає загально-ре\-ля\-ти\-віст\-ські ефекти при
поширенні електромагнітного випромінювання. У даній роботі
розглянуто ефекти, залежні від довжини хвилі, при (мікро)лінзуванні
протяжного гаусівського джерела на точковій масі за стандартних
припущень щодо некогерентності різних елементів джерела. Отримано
аналітичні вирази для спектра потужності мікролінзованого
випромінювання, що є ефективними за великого джерела. Коли центр
джерела, маса та спостерігач розташовані на одній прямій, знайдено
спектр потужності в замкненій формі через гіпергеометричну функцію.
У випадку загального розташування цю величину знайдено у формі ряду.
Отримано асимптотичні вирази для спектра потужності за великого
розміру джерела та за високих частот.}

\end{document}